\documentclass[aps,prl,twocolumn,superscriptaddress,amssymb,nofootinbib]{revtex4}

\usepackage{color}
\usepackage{epsfig}
\usepackage{graphicx}
\usepackage{epstopdf}

\begin{document}

\title{Light CP-odd Higgs and Small $\tan \beta$ Scenario in the MSSM and Beyond}

\author{Radovan Derm\' \i\v sek}
\email[]{dermisek@ias.edu}

\affiliation{School of Natural Sciences, Institute for Advanced Study, Princeton,
NJ 08540}



\date{June 5, 2008}

\begin{abstract}
We study the Higgs sector of supersymmetric models containing two Higgs doublets with a light MSSM-like CP odd Higgs, $m_A \lesssim 10$ GeV, and $\tan \beta \lesssim 2.5$. 
In this scenario
all of the Higgses resulting from two Higgs doublets: light and heavy CP even Higgses, $h$ and $H$, the CP odd Higgs, $A$, and the charged Higgs, $H^\pm$, could have been produced at LEP or the Tevatron, but would have escaped detection because they decay in modes that have not been searched for or the experiments are not sensitive to. Especially $H \to ZA$ and $H^\pm \to W^{\pm \star} A$ with $A \to c \bar c, \tau^+ \tau^-$
present an opportunity to discover some of the Higgses at LEP, the Tevatron and also at B factories. Typical  $\tau$- and $c$-rich 
decay products of all Higgses require modified strategies for their discovery at~the~LHC.
\end{abstract}


\maketitle







{\it Introduction:}
In our current understanding of particle physics 
all known elementary particles  acquire mass in the process of electroweak symmetry breaking.
In the standard model (SM) this is achieved by the Higgs mechanism: a complex scalar electroweak doublet
gets a vacuum expectation value, spontaneously breaks the electroweak symmetry, gives masses to $W$ and $Z$ bosons and leaves a scalar Higgs boson in the spectrum. 
The Higgs boson is the last missing piece of the standard model. 
In theories beyond the SM the Higgs sector is typically more complicated, e.g. in the minimal supersymmetric standard model (MSSM) there are two Higgs doublets which lead to five Higgs bosons in the spectrum:  light and heavy CP even Higgses, $h$ and $H$, the CP odd Higgs, $A$, and a pair of charged Higgs bosons, $H^\pm$; and there are many simple models with even more complicated Higgs sector.
Discovery of Higgs bosons and exploration of their properties is the key to understanding the electroweak symmetry breaking  and a major step in uncovering the ultimate theory of particle physics.

Since the searches for Higgs bosons rely on detection of their decay products, 
it is crucial to understand the way Higgs bosons decay. Although it is usually the case that there is one Higgs boson with properties (couplings to $W$ and $Z$ bosons) of the SM Higgs it is not necessarily true that such a Higgs decays in the way the SM Higgs does~\cite{Chang:2008cw}. A significant model dependence of decay modes applies to other Higgses as well. 

In this letter we would like to bring attention to the class of models with Higgs sectors that resemble the 
Higgs sector of the MSSM in the region with a light CP odd Higgs boson, $m_A \lesssim 10$ GeV, and $\tan \beta \lesssim 2.5$.
Although this region is ruled out in the MSSM, after careful review of experimental limits we argue that it is easy to make this region phenomenologically viable in simple extensions of the MSSM. 
We focus on features that  involve the two Higgs doublet part of a possible extension.  Perhaps the most interesting observation is 
that all the Higgses resulting from two Higgs doublets: $h$, $H$, $A$ and $H^\pm$ could have been produced already at LEP or the Tevatron, but would have escaped detection because they decay in modes that have not been searched for or the experiments are not sensitive to. We discuss several search modes that present an opportunity to discover some of the Higgses at both LEP and the Tevatron. These 
decay modes also require modified strategies for Higgs discovery at the LHC. In addition, the light CP odd Higgs might be within the reach of current B factories.


{\it MSSM at $m_A \ll m_Z$ and $\tan \beta \lesssim 2.5$:}
Let us start with the discussion of the Higgs sector of the MSSM~\cite{higgs_review}.
In the MSSM, the CP-even Higgs mass-squared matrix at the tree level 
can be written in terms of CP odd Higgs boson mass, $m_A$, the mass of the Z boson, $m_Z$, and the ratio of the vacuum expectation values of the two Higgs doublets,  $\tan \beta = v_u/v_d$. Radiative corrections can be approximated by the contribution to $H_u-H_u$ element of the Higgs mass-squared matrix. This contribution,  $\Delta$, 
is dominated by  stop loops and thus depends on
stop masses and  mixing in the
stop sector (it contains a $1/\sin^2 \beta$ factor). The Higgs mass eigenstates are obtained by an orthogonal transformation parameterized by $\alpha$.
The coupling squared of the lighter CP-even Higgs boson to $ZZ$ divided by the SM value is given by: 
\begin{equation}
C_{ZZh} = \frac{g^2_{ZZh}}{g^2_{ZZh_{SM}}} =  \sin^2 (\beta - \alpha).
\end{equation}
The two CP even Higgs bosons share the SM Higgs coupling to Z and thus $C_{ZZH} = \cos^2 (\beta - \alpha)$.
For Higgs searches also the couplings of $h$ and $H$ to $ZA$ states 
are important, and their values squared (appropriately rescaled)
are given as: $C_{ZAh} = \cos^2  (\beta - \alpha)$ and $C_{ZAH} = \sin^2  (\beta - \alpha)$. Note 
that the couplings are complementary which is important for Higgs searches.

The Higgs spectrum is particularly simple in two distinct regimes. In the decoupling regime, $m_A \gg m_Z$, masses of light and heavy CP even Higgses are given by
\begin{eqnarray}
m_h^2 &\simeq & m_Z^2 \cos^2 2 \beta + \Delta \sin^2 \beta , \\
m_H^2 &\simeq & m_A^2 + m_Z^2 \sin^2 2 \beta + \Delta \cos^2 \beta \simeq m_A^2 .
\label{eq:mH2decoupled}
\end{eqnarray}
The light CP even Higgs boson is SM-like in its couplings to $ZZ$ and $WW$ for any $\tan \beta \geq 1$:
\begin{equation}
C_{ZZh} \simeq 1 - \frac{m_Z^4}{4 m_A^4} \sin^2 4\beta \simeq 1.
\end{equation}

In the opposite limit,  characterized by $m_A \ll m_Z$, 
neglecting  radiative corrections we have:
\begin{eqnarray}
m_h^2 &\simeq& m_A^2 \cos^2 2 \beta ,
\label{eq:mh2_mAlessmZ} \\
m_H^2 &\simeq& m_Z^2(1+ \frac{m_A^2}{m_Z^2}\sin^2 2 \beta),
\end{eqnarray}
and the $ZZH$ coupling is given by:
\begin{equation}
C_{ZZH} \simeq \cos^2 2\beta \left( 1 - \frac{m_A^2}{m_Z^2} \sin^2 2\beta \right).
\label{eq:ZZH_mAlessmZ}
\end{equation}
Formulas including radiative corrections are somewhat long and not particularly revealing in this limit. Note however, that for  $\tan \beta > few$ the radiative corrections contribute mostly to the Heavy CP even Higgs which plays the role of the SM Higgs, $C_{ZZH} \simeq 1$.
This regime of the parameter space does not receive much attention because it is beyond any doubt ruled out by Higgs searches at LEP.
The mass of the light CP even Higgs cannot be raised by radiative corrections and pairs of $h A$ would be copiously produced since $g_{ZAh} \simeq 1$. In addition, light $h$ and $A$ would significantly contribute to the Z width and thus are ruled out 
by the Z-pole measurements~\cite{:2004qh}.

As $\tan \beta $ approaches 1 for $m_A \ll m_Z$ the situation described above dramatically changes.\footnote{This region of the parameter space has an interesting physical meaning. A light CP odd Higgs is a signal of an approximate
global $U(1)_R$ symmetry of the Higgs potential which allows the $\mu$-term in the superpotential, $W \supset \mu H_u H_d$, but forbids the corresponding $B_\mu$-term in the soft SUSY breaking lagrangian. This symmetry is broken by gaugino masses and thus a small $B_\mu$-term is generated by radiative corrections that lift the mass of the CP odd Higgs, $m_A = 2 B_\mu  \sin 2 \beta$. The small $\tan \beta$ on the other hand signals that the top Yukawa couplings becomes non-perturbative close to the grand unification (GUT) scale. The exact value of $\tan \beta$ consistent with perturbativity all the way to the GUT scale depends on superpartner masses through threshold corrections to the top Yukawa coupling, and it is about $\tan \beta \gtrsim 1.2$. However, adding extra vector-like complete SU(5) matter multiplets at the TeV scale, e. g. parts of the sector that mediates SUSY breaking (messengers) or present for no particular reason, does not affect the unification of gauge couplings while it slows down the running of the top Yukawa coupling~\cite{Masip:1998jc, Barbieri:2007tu} and even $\tan \beta \simeq 1$ can be consistent with perturbative unification of gauge couplings.} 
The light CP even Higgs boson becomes SM-like, $C_{ZZh} \simeq 1$, since $C_{ZZH} \simeq 0$, see Eq.~(\ref{eq:ZZH_mAlessmZ}), and although
it is massless at the tree level~(\ref{eq:mh2_mAlessmZ}), it will receive a contribution from superpartners and the tree level relation between the light CP even and CP odd Higgses, $m_h < m_A$ is typically not valid. Even for modest superpartner masses the light CP even Higgs boson will be heavier than $2m_A$ and thus $h \to AA$ decay mode is open and generically dominant. For small $\tan \beta$ the width of $A$ is shared between $\tau^+ \tau^-$ and $c \bar c$ for $m_A < 2m_b$ and thus the width of $h$ is spread over several different final states, $4\tau$, $4c$, $2\tau 2c$ and highly suppressed $b \bar b$ and thus the LEP limits in each channel separately are highly weakened.
Since $h$ is SM-like,  $e+ e^- \to hA$ is highly suppressed and the limits from the Z width measurements can be easily satisfied even for  $m_h + m_A < m_Z$. 
In addition, we will see that decay modes of the heavy CP even Higgs (that also turns out to be within the reach of LEP) are modified in this region and even the charged Higgs boson can be below LEP or Tevatron limits due to decay modes that have not been searched for.

For $m_A \ll m_Z$ and $\tan \beta \simeq 1$ including radiative corrections we find:
\begin{eqnarray}
m_h^2 &\simeq& \Delta/2, \\
m_H^2 &\simeq& m_Z^2 + m_A^2 + \Delta/2. 
\end{eqnarray}
As already mentioned, $h$ is SM-like and thus $C_{ZZh} = C_{ZAH} \simeq 1$ and  $C_{ZZH} = C_{ZAh} \simeq 0$.
For  $\tan \beta \simeq 2.5$ both Higgses equally share the coupling to $Z$, $C_{ZZh} \simeq C_{ZZH} \simeq 0.5$.
The mass of the charged Higgs is given as,
\begin{eqnarray}
m_{H^\pm}^2 &=& m_W^2 + m_A^2 - \Delta^\prime \; \simeq \; m_W^2, 
\end{eqnarray}
where $\Delta^\prime$ represents radiative correction which is typically not significant  (it is positive and has a tendency to decrease the mass of the charged Higgs). 

Using {\it FeynHiggs2.6.3}~\cite{Heinemeyer:1998yj} (with $m_t = 172.6$ GeV) for $\tan \beta = 1.01$, $\mu = 100$ GeV, $m_A = 8$ GeV and varying soft susy breaking scalar and gaugino masses between 300 GeV and 1 TeV and mixing in the stop sector, $X_t/m_{\tilde t}$, between 0 and -2, we typically find: $m_h \simeq 38 - 56$ GeV with $g_{ZZh}/g_{ZZh_{SM}} \simeq 0.84 - 0.97$, $m_H \simeq 108 - 150$ GeV and  $m_{H^\pm} \simeq 78 - 80$ GeV. The dominant branching ratios of the light CP even Higgs are typically:
\begin{equation}
B (h \to A A, \; b \bar b)  \; \simeq \;  90 \%,  \; 10 \% 
\label{eq:Bh}
\end{equation}
with 
\begin{equation}
B (A \to \tau^+ \tau^-, \; c \bar c, \; gg )  \; \simeq \;  50 \%, \; 40 \%, \; 10\% ,
\label{eq:BA}
\end{equation}
for $2m_\tau \lesssim m_A \lesssim 10$ GeV.
Branching ratios of the Heavy CP even Higgs vary with SUSY spectrum. 
For 1 TeV SUSY and $X_t/m_{\tilde t} = 0$ we find:
\begin{equation}
B (H \to ZA, \; A A, \; hh, \; b \bar b)  \; \simeq \;  37 \%, \; 34 \%, \; 28 \% , \; 0.4 \%
\label{eq:BH}
\end{equation}
(similar branching ratios apply for 300 GeV SUSY with $X_t/m_{\tilde t} = -2$).
Finally, the dominant branching ratios of the charged Higgs are:
\begin{equation}
B (H^+ \to W^{+ \star } A, \; \tau^+ \nu, \; c \bar s)  \; \simeq \;  70 \%, \; 20 \%, \; 10 \% . 
\label{eq:BHpm}
\end{equation}
For discussion of experimental constraints let us also include branching ratios of the top quark:
\begin{equation}
B(t \to H^+ b) \simeq 40 \%, \quad B(t \to W^+ b) \simeq 60 \% .
\label{eq:Bt}
\end{equation}
These results (except (\ref{eq:BH})) are not 
very sensitive to superpartner masses nor 
the mass of the CP odd Higgs as far as $m_A < 2m_b$. Increasing $\tan \beta$ to $2.5$ 
only the following branching ratios significantly change: $B (A \to \tau^+ \tau^-, gg )  \simeq  90 \%,  10\%$, $B (H^+ \to W^{+ \star } A, \tau^+ \nu)  \simeq  35 \%, 65 \%$ and $B(t \to H^+ b, W^+ b) \simeq 10 \%, 90 \%$.


In spite of all Higgses being within the reach of LEP we will see that only the light CP even Higgs is
significantly constrained and actually in the MSSM ruled out by LEP data. For  limits on various search  modes and complete list of references see Ref.~\cite{Schael:2006cr}. 


{\it Light CP even Higgs:}
One of the strongest limits on the CP even Higgs boson comes from the search for $h \to b \bar b$.
The limit for  $m_h \lesssim 60$ GeV is  $C_{ZZh} B(h \to b \bar b) \lesssim 0.04$. In order to satisfy this limit we need $B (h \to A A) \gtrsim 96 \%$. This is not impossible, but it is not generically (in large regions of SUSY parameter space) satisfied~(\ref{eq:Bh}). 
With generic $B (h \to b \bar b) \simeq 10 \%$ we need $m_h \gtrsim 85$ GeV  or somewhat reduced $C_{ZZh}$.
Note also that combined $h \to b \bar b$ and  $h \to AA \to 4b$ limit requires $m_h > 110$ GeV for $C_{ZZh} \simeq 1$. Thus $m_A < 10$ GeV is favored. For $m_A < 10$ GeV the limits on the dominant decay mode $h \to AA$ are not so strong  since it is spread over several final states: $h \to AA \to 2\tau 2c, 4\tau, 4c, \dots$ with about 36\%, 23\%, 14\%, $\dots$ branching ratios. These limits can be satisfied for any $m_h$ provided $m_A\gtrsim 9$ GeV with limits weakening for heavier $h$ and completely expiring for  $m_h \gtrsim 86$ GeV~\cite{Jack_TeV4LHC}. 

The limits from $e^+ e^- \to hA$ 
are typically comfortably satisfied for masses and branching ratios of interest as a consequence of a very small $C_{ZAh}$. The strongest limit comes from $hA \to AAA \to 6 \tau$ which for $m_A \simeq 10$ GeV and $m_h = 40 - 60$ GeV requires $C_{ZAh} \lesssim 0.07$ and this limit runs out for  $m_h \simeq 65$ GeV. A comparable 
constraint on $C_{ZAh}$ comes from the $Z$-width measurement for $m_h = 40$ GeV and it becomes weaker for larger $m_h$.

We see that this scenario can avoid all decay-mode specific  limits in the region of SUSY parameter space that leads to $B (h \to A A) > 96 \%$ and $C_{ZZh} \gtrsim 0.93$ and we will se  later that there are no other relevant constraints from searches for $H$ and $H^\pm$.  However, the decay-mode independent search from OPAL~\cite{Abbiendi:2002qp} sets the limit on the Higgs mass by looking only for reconstructed Z boson decaying leptonically and excludes $m_h < 82$ GeV for $C_{ZZh} = 1$.  
 
Taking the decay-mode independent limit at face value\footnote{We would like to make two comments however. First of all, this search was not expected to be sensitive to $m_h \gtrsim 55$ GeV and much stronger limits than expected were set due to a significant deficit ($\sim 2\sigma$) of background events for a large range of $m_h$ between 60 an 80 GeV  (deficit of background events is also responsible for much stronger limits on $C_{ZZh} B(h \to b \bar b)$ in the same region). Second of all, the efficiencies for signal events to pass the selection cuts are estimated assuming the Higgs decays into two body final states. 
Clearly, the efficiencies for Higgs decaying into four body final states
$4\tau, 4c, 2\tau 2c, \dots$ (that further decay into states typically containing $e^\pm$, $\mu^\pm$ either from $\tau$ or semi-leptonic $c$ decays which might make the Z reconstruction less efficient) are smaller and the exclusion limits would be weaker.} our scenario is ruled out in the MSSM, since $m_h$ cannot be pushed above 82 GeV with radiative corrections. There are however various ways to increase the mass of the SM-like Higgs boson in extensions of the MSSM. A simple possibility is to consider singlet extensions of the MSSM containing $\lambda S H_u H_d$ term in the superpotential. It is known that this term itself contributes $\lambda^2 v^2 sin^2 2 \beta$, where v = 174 GeV, to the mass squared of the CP even Higgs~\cite{Ellis:1988er} and thus can easily push the Higgs mass above $82$ GeV. Note this contribution is maximized for $\tan \beta \simeq 1$. Singlet extensions can also alter the couplings of the Higgses to $Z$ and $W$ through mixing~\cite{mixed} or provide new Higgs decay modes~\cite{4tau},~\cite{Chang:2008cw}. Considering this possibility would lead to very model dependent 
predictions and thus in this paper we assume that a possible extension does not significantly alter the two Higgs doublet part of the Higgs sector besides increasing the Higgs mass above the decay-mode independent limit.


{\it Heavy CP even Higgs:}
The heavy CP even Higgs is too heavy to be produced in association with $Z$ at LEP. In addition, $C_{ZZH}$ is small. However, since $C_{ZAH} \simeq 1$ and A is light, pairs of $AH$ would be produced at LEP. The strongest limit comes from $HA \to b \bar b \tau^+ \tau^-$ which is however comfortably satisfied due to small $B(H\to b \bar b)$.
The limits on $AH \to Ahh$ mode are not constraining since $h$ decays dominantly into $AA$. This mode 
might not be open in extensions of the MSSM which increase the mass of $h$ in which case $ZA$, $AA$ and $b\bar b$ will share the width of $H$.  The searches in various final states of $AH \to AAA$ are either not sensitive to or were not done in the range of masses typical in our scenario. 
{\it The dominant $H \to ZA$ mode has never been searched for!}
Interestingly, $e^+ e^- \to H A \to (ZA) A$ events could be mistaken for $e^+ e^- \to ZH \to Z (AA)$ or simply $Z +jets$ for which searches have been done. Of course the interpretation (the reconstructed Higgs mass) would be wrong and this might be responsible for various local excesses of events and small excesses over large range of reconstructed Higgs masses in flavor independent searches, see e.g. Refs.~\cite{Achard:2003ty, Abdallah:2004bb}.


{\it Charged Higgs:}
Searches for pair produced charged Higgs bosons were performed by LEP 
collaborations~\cite{:2001xy, Achard:2003gt, Heister:2002ev, Abdallah:2003wd}.
Three different final states, $\tau^+ \nu \tau^- \bar \nu$, $c \bar s \bar c s$ and
$c \bar s \tau^- \bar \nu$ were considered
and lower limits were set on the mass $m_{H^\pm}$ 
as a function of the branching ratio $B(H^+ \to \tau^+ \nu)$, assuming 
$B(H^+ \to \tau^+ \nu) + B(H^+ \to c \bar s)=1$.
In addition, DELPHI considered a possibility $H^+ \to W^{+\star} A$ which is important if the CP
odd Higgs boson is not too heavy~\cite{Akeroyd:2002hh} and limits were obtained under the assumption that $A$  is heavy
enough to decay into $b \bar b$~\cite{Abdallah:2003wd}.

The strongest limits are set by ALEPH~\cite{Heister:2002ev}. Assuming $B(H^+ \to \tau^+ \nu) + B(H^+ \to c \bar s)=1$, charged Higgs bosons with mass below 79.3 GeV are excluded at 95\% C.L., independent of $B(H^+ \to \tau^+ \nu)$.
Somewhat lower limits have been obtained by DELPHI~\cite{Abdallah:2003wd} and L3~\cite{Achard:2003gt} collaborations due to local excesses of events.

In the scenario discussed above the charged Higgs decays dominantly into $W^\star A$ with $A \to c \bar c$ or $\tau^+ \tau^-$~(\ref{eq:BHpm}). LEP limits thus apply to the remaining branching ratios and can be comfortably satisfied for  $m_{H^\pm} \gtrsim 75$ GeV.

At the Tevatron the charged Higgs is searched for in the decay of the top quark.
If kinematically allowed, the top quark can decay to $H^+ b$, competing with the standard model decay $W^+ b$. The strongest limits come from CDF~\cite{Abulencia:2005jd} from
data samples corresponding to  an integrated luminosity of 193 ${\rm pb}^{-1}$.
It is assumed that $H^+$ can decay only to $\tau^+ \nu$, $c\bar s$, $t^\star \bar b$ or $W^+ A$ with $A \to b \bar b$. 

If charged Higgs decays exclusively to $\tau^+ \nu$, the $B(t \to H^+ b)$ is constrained to be less than 0.4 at 95 \% C.L. 
For MSSM benchmark scenarios, assuming $H^+ \to \tau^+ \nu$ or $H^+ \to c \bar s$ only, stronger limits than at LEP are set for $\tan \beta \lesssim 1.3$ on the mass of the charged Higgs. For  $\tan \beta \lesssim 1$ the limit is $m_{H^\pm} \gtrsim 100$ GeV. 
If no assumption is made on the charged Higgs decay (but still allowing only those that were searched for) the  $B(t \to H^+ b)$ is constrained to be less than $\sim 0.8$ for $m_{H^\pm} \simeq 80$ GeV.

For  $H^\pm \to W^{\pm \star} A$ with $A \to c \bar c$ or $\tau^+ \tau^-$,
{\it the decay modes that were not search for}, and in addition modes that can easily mimic $W$ decay modes, especially the dominant hadronic mode, it is reasonable to expect that the limits would be even weaker. In our scenario $B(t \to H^+ b) \lesssim 40 \%$, see Eq.~(\ref{eq:Bt}), and thus the Tevatron  does not place stronger limits than LEP. 



{\it CP odd Higgs:}
The light CP odd Higgs might be within the reach of current B factories where it can be produced  
in Upsilon decays, $\Upsilon \to A \gamma$. 
This was recently suggested in the framework of the
next-to-minimal supersymmetric model (NMSSM) with a light CP odd Higgs boson being mostly the singlet~\cite{Dermisek:2006py} (in our scenario $A$ is doublet-like) and it overlaps with searches for lepton non-universality in $\Upsilon$ decays~\cite{Sanchis-Lozano}. It is advantageous to look for a light CP odd Higgs in $\Upsilon (1S)$, $(2S)$ and $(3S)$ since these states cannot decay to B mesons and thus the $A \gamma$ branching ratio is enhanced.
Predictions for the branching ratio $B(\Upsilon \to A \gamma)$ for $\tan \beta = 1$ can be readily (although only approximately) obtained from the results of Ref.~\cite{Dermisek:2006py} taking $\tan \beta \cos \theta_A \simeq 1$ ($\cos \theta_A$ is the doublet component  which is 1 for MSSM-like CP odd Higgs).
We find, that the  $B(\Upsilon (1 S) \to A \gamma) $ ranges between $5 \times 10^{-5}$ for $m_A \simeq 2 m_\tau$ and  $10^{-7}$ for $m_A \simeq 9.2$ GeV. 

CLEO recently reported results from $\Upsilon (2S) \to \pi^+ \pi^- \Upsilon (1S) \to \pi^+ \pi^- A \gamma$~\cite{Kreinick:2007gh} and also preliminary results from  $\Upsilon (1S)$ data~\cite{CLEOprelim}.
 The limits, assuming $B(A \to \tau^+ \tau^-)= 1$, range between $7 \times 10^{-5}$ and $8 \times 10^{-6}$ depending on the mass of the CP odd Higgs. For large $\tan \beta$ these limits rule out a significant part of the parameter space. For small $\tan \beta$ the $B(A \to \tau^+ \tau^-)$ is reduced and thus these limits
are less constraining.
The BaBar results from recently taken $\Upsilon (3S)$ and $\Upsilon (2S)$ data assuming $B(A \to \tau^+ \tau^-) = 1$ should be available soon~\cite{BaBar}.



{\it Discussion:}
Supersymmetric models with Higgs sectors that resemble the 
Higgs sector of the MSSM in the region with light CP odd Higgs boson, $m_A \lesssim 10$ GeV, and $\tan \beta \lesssim 2.5$ can have SM-like CP even Higgs boson as light as 82 GeV (the decay-mode independent limit) dominantly decaying into $h \to AA \to 2\tau 2c, 4\tau, 4c$.
This feature is similar to the scenario with a light singlet-like CP-odd Higgs in the NMSSM~\cite{4tau}.   
In addition, this scenario predicts that the CP odd and the heavy CP even Higgses were also produced at LEP in
 $e^+ e^- \to HA$, and could be discovered searching for the dominat decay mode $H \to Z A$. 
Furthermore, up to $\sim 40 \%$  of top quarks produced at the Tevatron could have decayed into charged Higgs and the $b$ quark with $H^\pm \to W^{\pm \star} A$ and $A \to c \bar c$ or $\tau^+ \tau^-$.
These decay modes of $H$ and $H^\pm$ have never been searched for and thus provide an opportunity to discover some of the Higgses in LEP data or at the Tevatron. The search for $H^\pm$ including these modes is especially desirable at the Tevatron with currently available large data sample. The need for these searches is further amplified by the fact that the charged Higgs with properties emerging in this scenario could explain the $2.8 \sigma$ deviation from lepton universality in $W$ decays measured at LEP~\cite{Dermisek:2008dq}.

The CP odd Higgs could also be discovered at B factories. It would be desirable to include $A \to c \bar c$ and combine searches assuming 
$B(A \to \tau^+ \tau^-)  + B(A \to c \bar c) = 1$ or, optimally, allowing a small $B(A\to gg)$.

Since dominant decay modes of $h$, $H$, $A$ and $H^\pm$ lead to $\tau$- and $c$-rich final states (this is very different from usual scenarios where Higgs decays are dominated by $b$ final states) modified strategies for their discovery at the LHC are needed.

Although this scenario is ruled out in the MSSM we have argued that it can be easily viable in simple extensions. For example,
in the NMSSM the scenario with a light doublet-like 
CP odd Higgs and small $\tan \beta$ is viable and has all the features (Higgs mass ranges and decay modes) of the MSSM in this 
limit~\cite{NMSSM_small_tb}. It should be stressed however that this scenario is 
not limited to singlet extensions of the MSSM and it would be viable in many models 
beyond the MSSM that increase 
the mass of the SM-like Higgs boson.

\acknowledgments

\vspace{0.2cm}
I would like to thank  A. Akeroyd, J. Gunion,  P. Langacker, B. McElrath, V. Rychkov and D. Shih  for discussions.
RD is supported by the U.S. Department of Energy, grant DE-FG02-90ER40542. 

\vspace{0.2cm}



\end{document}